\documentclass[aps,pre,showpacs,twocolumn,showkeys,amssymb,amsmath,superscriptaddress]{revtex4-1}
\usepackage[colorlinks=true,linkcolor=red]{hyperref}
\usepackage{graphicx}
\usepackage{amsmath}
\usepackage{amsfonts}
\usepackage{amssymb}
\usepackage{color}
\usepackage[sort&compress]{natbib}

\newcommand{\be}{\begin{equation}}
\newcommand{\ee}{\end{equation}}
\newcommand{\bea}{\begin{eqnarray}}
\newcommand{\eea}{\end{eqnarray}}

\newcommand{\nn}{\nonumber}

\newcommand{\vs}{\vec{S}}
\newcommand{\vm}{\vec{m}}
\newcommand{\vh}{\vec{h}}
\newcommand{\vy}{\vec{y}}
\newcommand{\vn}{\vec{n}}
\newcommand{\vchi}{\vec{\chi}}
\newcommand{\vex}{\hat{e}_x}
\newcommand{\mch}{\ensuremath{\mathcal{H}}}

\newcommand{\tpr}{\ensuremath{\textrm{PR}}}

\begin{document}

\title{Avalanches and hysteresis in frustrated superconductors and XY-spin-glasses}

\author{Auditya Sharma}
\affiliation{International Institute of Physics - Federal University of Rio Grande do Norte, Natal, RN, Brazil}
\affiliation{Tel Aviv University, Tel Aviv, Israel}
\author{Alexei Andreanov}
\affiliation{The Abdus Salam ICTP, Strada Costiera 11, I-34151 Trieste, Italy}
\affiliation{Max Planck Institute for Physics of Complex Systems\\
N\"othnitzer Str. 38, 01187 Dresden, Germany}
\author{Markus M\"uller}
\affiliation{The Abdus Salam ICTP, Strada Costiera 11, I-34151 Trieste, Italy}
\date{\today}

\begin{abstract}
	We study avalanches along the hysteresis loop of long-range interacting spin-glasses with continuous XY-symmetry - which serves as a toy model of granular superconductors with long-range and frustrated Josephson couplings. We identify sudden jumps in the $T=0$ configurations of the XY-phases, as an external field is increased. They are initiated by the softest mode of the inverse susceptibility matrix becoming unstable, which induces an avalanche of phase updates (or spin alignments). We analyze the statistics of these events, and study the correlation between the non-linear avalanches and the soft mode that initiates them. We find that the avalanches follow the directions of a small fraction of the softest modes of the inverse susceptibility matrix, similarly as was found in avalanches in jammed systems. In contrast to the similar Ising spin-glass (Sherrington-Kirkpatrick) studied previously, we find that avalanches are not distributed with a scale-free power law, but rather have a typical size which scales with the system size. We also observe that the Hessians of the spin-glass minima are not part of standard random matrix ensembles as the lowest eigenvector has a fractal support.
\end{abstract}

\pacs{
75.10.Hk	
75.50.Lk	
75.60.Ej	
}

\maketitle

\section{Introduction}


Hysteresis is a ubiquitous phenomenon, encountered in a wide range of disordered systems which can be trapped in long-lived metastable configurations. In a typical experiment, a control parameter (e.g., an external field) is varied cyclically, while a physical observable (e.g., the magnetization) is tracked. In the presence of metastable states, the path taken along the forward direction is usually different from that on the reverse direction, displaying a dependence  on the history, and thus memory effects~\cite{bertotti1998hysteresis}.

In the presence of strong randomness in magnets, the polarization proceeds in mesoscopic bursts, where at certain specific values of the applied field the change of orientation of a portion of the system triggers a large rearrangement, referred to as an avalanche~\cite{sethna1993hysteresis,dhar1997zero,pazmandi1999self,sabhapandit2000distribution,le2010avalanches}. In ferromagnets this phenomenon is well-known as Barkhausen noise. Such avalanches have been the subject of considerable interest in recent years~\cite{sethna1993hysteresis,vives1995universality,dhar1997zero,pazmandi1999self,ledoussal2009driven,ledoussal2009size, ledoussal2009statistics,andresen2013self,ispanovity2013avalanches}. Under certain circumstances, the distribution of avalanches may become critical, characterized by a scale-free power law, cut off only by a scale that diverges with the system size. This happens for example at the depinning threshold of pinned elastic interfaces (domain walls), where the criticality of avalanche distributions reflects the dynamical criticality of the depinning transition~\cite{tanguy1998probamat,paczuski1996avalanche,rosso2009avalanche}. The latter governs a wide variety of phenomena like earthquakes, domain wall motion in magnets, crackling noise, sandpile models etc~\cite{emig2000exact,fisher1998collective,bonami2008crackling,altshuler2004vortex,bak1987self,dhar1999abelian,vives1995universality,field1995superconducting} .

However, in simple toy models of random ferromagnets, such as the random field Ising model, criticality usually requires  fine-tuning, both of the disorder strength as well as of the external field~\cite{dahmen1996hysteresis}. In more realistic descriptions of experiments, the negative feedback from demagnetization fields can, however, ensure the existence of a parameter window in which critical response along the hysteresis loop is observed~\cite{colaiori2008exactly}.

Interestingly, in the Sherrington-Kirkpatrick Ising spin-glass, a  frustrated magnet with fully connected interactions, such criticality was numerically observed along the entire hysteresis loop, without requiring any fine-tuning~\cite{pazmandi1999self}. A very similar phenomenology of system-spanning avalanches that require no fine-tuning was  found in the avalanche dynamics of long-range interacting 2d dislocation systems~\cite{ispanovity2013avalanches}. The criticality found in the SK model was interpreted as a manifestation of the self-organized criticality of the relevant out-of-equilibrium configurations visited in the spin-glass phase~\cite{pazmandi1999self}. A calculation of the power-law distributed \emph{equilibrium} avalanches in the same system suggested that there might indeed be a close link between the well-known marginal stability of the spin-glass phase (at equilibrium), and the observed scale-free avalanches out-of-equilibrium~\cite{le2010avalanches,ledoussal2012equilibrium}.

On the other hand, a recent study of short-range spin-glass models on random graphs has shown that avalanches in such systems do not follow a scale-free distribution, in spite of their equilibrium being expected to be marginally stable. That study suggested that is the long range of the interactions in the SK model, rather than its thermodynamic marginality, that plays the crucial role in ensuring scale-free avalanches~\cite{andresen2013self}. In the physically interesting intermediate case of power law interactions, such as unscreened Coulomb interactions which decay as $1/r$ with distance, it apears that whether or not scale-free avalanches are observed in the hysteresis depends on the constraints imposed on the dynamics~\cite{palassini2012elementary,andresen2013self}.

\subsection{Ising versus vector spins}


In essentially all of the above examples, the ordering degrees of freedom have a discrete, Ising-like character. 
In the present paper we instead investigate avalanches in a system with continuous degrees of freedom, and contrast its phenomenology with that of Ising systems. In particular, we focus on vector spin-glasses with $m=2$ components (XY-spins). Those can be considered as toy models describing granular superconductors with Josephson couplings, that are frustrated by the presence of an external flux. 
Since Josephson couplings decay only as a power law in space, we consider here the case of infinite range, SK-type interactions, and focus on the effect brought about by the spin rotation symmetry on the phenomenology of the hysteresis, and in particular the statistics of avalanches, as compared to the Ising case.
A particular realization of such a system with very long ranged couplings is the ``superconducting hay'' proposed and studied in Refs.~\onlinecite{ebner1985diamagnetic,vinokur1987system,feigelman1995theory},  an assembly of needle-shaped, superconducting islands, each of which having many crossing junctions with other needles.
  
Vector spin-glasses exhibit a variety of new features as compared to their Ising counterparts, both in and out of equilibrium. In contrast to the Ising case, for short-range systems, the existence of a spin-glass transition at finite $T$ has been debated for a long time, as well as the role of chirality~\cite{villain1977two,kawamura1987chiral,lee2007large,sharma2011phase}. In the presence of magnetic fields, one has to distinguish uniform and random orientations of the fields. Mean-field theory in a uniform field predicts the Gabay-Toulouse transition line, where the transverse components undergo freezing and spontaneously break the symmetry of rotations around the axis of the external field~\cite{gabay1981coexistence}. In the presence of randomly oriented external fields, there is no symmetry left to be broken, but a phase transition persists along the famous Almeida-Thouless line~\cite{de1978stability} for arbitrary $m$-component vector spin glasses, as was shown within mean-field theory in Ref.~\onlinecite{sharma2010almeida}. Whether or not the Almeida-Thouless line exists for short-range models, one of the crucial questions in the theory of spin glasses, is also an interesting open question for vector models~\cite{sharma2011almeida}.

The equilibrium properties of infinite range models, whose off-equilibrium counterpart we will study below, have been discussed in Refs.~\onlinecite{palmer1979internal,gabay1981coexistence,cragg1982instabilities,gabay1982symmetry,moore1982critical}. However, unlike for the SK model, a complete understanding of the $T\to 0$ limit of the replica symmetry breaking (RSB) solution is still lacking. The latter would be needed to analytically describe equilibrium avalanches (or shocks) in these systems. Below, we  focus instead on the out-of-equilibrium properties and avalanches along the hysteresis loop. However, if we \emph{assume} a close similarity between equilibrium and dynamic response (as was found in the SK model), we may infer conjectures about the structure of the overlap function $P(q)$, based on the avalanche distribution observed in the dynamics.  

Hysteresis in finite-dimensional XY- and Heisenberg ferromagnets with random field disorder was found to generically exhibit similar critical behavior as random field Ising magnets~\cite{silveira1999critical} (even though, upon tuning an extra parameter, a different universality class of critical avalanches was observed). In contrast, the case of long-range frustrated spin-glasses with continuous symmetry brings about new aspects of phenomenology as compared to the Ising counterpart. Like in the long-range Ising spin-glass, one expects very large avalanches to occur with finite probability. However, we will find that in the XY-glass most avalanches have a typical size which is set by the system size. The probability of very small avalanches is found to be rather negligible, and it grows as a power law with increasing avalanche size. Interestingly, unlike in the Ising case the distribution of avalanche sizes is thus not scale-free.

Another interesting aspect of the continuous spin symmetry is that, in contrast to Ising systems, the linear response within a metastable state remains non-trivial, even at $T=0$, since a change in  external field induces a smooth change of all angles, whereas Ising spins start flipping only when the local field of the least stable spin changes sign. It is thus  interesting to study the linear modes which dominate the susceptibility and analyze their relation with the non-linear avalanches that are triggered when the softest  of those modes becomes unstable. We emphasize also that, unlike in the Ising case, where an avalanche is triggered by a single spin-flip, avalanches in XY-systems are induced by the instability of a \emph{collective} mode that typically involves many spins. 


The above features are in fact analogous to avalanche phenomena in other glassy systems with continuous degrees of freedom. In particular, it is interesting to compare them with  jammed soft matter systems, which exhibit jumps in their evolution under applied shear stress~\cite{lin2014on} or relaxation~\cite{brito2007heterogeneous}. In those systems it was found that the non-linear jump events are strongly correlated to the softest modes of a Hessian matrix governing the linear fluctuations around the initial metastable state~\cite{brito2007heterogeneous}.


The remainder of this paper is organized as follows: in Sec.~\ref{sec:model}, we define the infinite range XY-glass, and describe the dynamics studied at $T=0$, as well as the observables and analytical criteria that determine jumps. Sec.~\ref{sec:results} analyzes the statistics of jump events, as obtained from numerical simulations of the XY-spin-glass. Sec.~\ref{sec:conclusions} summarizes the results and contrasts them with other systems exhibiting avalanches. 
In Appendix~\ref{app:TAP}, the results of section~\ref{sec:model} are rederived as the $T\to 0$ limit of a finite-$T$ calculation based on Thouless-Anderson-Palmer equations~\cite{thouless1977solution,plefka1982convergence}.

\section{Fully connected XY-glass}
\label{sec:model}

We consider a fully connected system of XY-spins,  i.e., the $2$-component version of the Sherrington-Kirkpatrick~\cite{sherrington1975solvable} model for spin-glasses, with Hamiltonian
\begin{gather}
	\label{eqn:ham}
	\mch = - \frac{1}{2}\sum_{ij}J_{ij}\vs_i\cdot\vs_j - \sum_i\vh_\textrm{ext}\cdot\vs_i .
\end{gather}
Here, $\vs_i =(S^x_i,S^y_i)$ are classical 2-component vectors of unit length: $\vs_i^2 = 1$ in the XY-plane, and $\vh_\textrm{ext}$ is a homogeneous external magnetic field. For convenience, we choose it to always point in the $x$-direction, 
\begin{gather}
	\vh_\textrm{ext} = H \vex.
\end{gather} 
The random bonds $J_{ij}$ are independently drawn from a Gaussian distribution, 
\begin{gather}
	\label{eqn:J_dist}
	P(J_{ij}) = \frac{1}{\sqrt{2\pi J^2/N}} e^{-N\,J_{ij}^{2}/J^{2}},
\end{gather}
where $N$ is the number of spins. Below we fix the energy units by setting $J = 1$.

This Hamiltonian also describes the classical limit of large superconducting islands with a well developed order parameter. They are characterized by a phase $\phi_i$, whose quantum dynamics we neglect, assuming a very small charging energy. In this realization, the interactions $J_{ij}$ between the islands arise due to Josephson couplings. In specific geometries, where the islands are needle-like structures that come close to many others without touching them (being spaced by insulating layers), and by applying a frustrating magnetic flux, such couplings can be both very long-range and random in sign~\cite{ebner1985diamagnetic,vinokur1987system,giura1990effects,giura1992flux,gartstein1994upper,feigelman1995theory,spivak2008theory},  which motivates the simplified toy model Eq~\eqref{eqn:ham}. Using the parametrization of spins by their angle in the plane, $\vs_i=(\cos\phi_i,\sin\phi_i)$, as measured from the positive $x$-axis, we can rewrite the Hamiltonian as: 
\begin{gather}
	\label{eqn:ham_phi}
	\mch = - \frac{1}{2}\sum_{ij}J_{ij}\cos(\phi_i-\phi_j) - H\sum_i\cos(\phi_i).
\end{gather}
By solving the adiabatic evolution under slow variations of magnetic field, Feigelman and Ioffe~\cite{feigelman1995theory} have shown that in such frustrated ``superconducting hay'', catastrophic events take place when a bias $H$ is applied to the angles (e.g., by Josephson-coupling all islands to a big superconductor, and homogeneously increasing  the coupling strength to this island). Such catastrophic events occur even when the evolution of this external bias is adiabatic. As we will discuss below they correspond to `phase avalanches', analogous to magnetization avalanches in Barkhausen noise.

\subsection{Polarization process at $T=0$}

We analyze this phenomenon adopting the XY-spin language for simplicity. We follow locally stable states, as the external field $H$ is varied slowly, and investigate the sudden jump-like events which occur as the frustrated system is more and more polarized. In the analogous situation in long-range Ising spin-glasses, it is known that the magnetization response occurs in avalanche-like steps of mesoscopic size~\cite{pazmandi1999self}. An analogue of this must  also be expected in the case of continuous spin symmetry. However, there is a significant difference. In the present case, the instabilities which induce avalanche-like events in the rearrangement of the XY-angles  are collective soft modes where a large number of spins moves coherently, whereas the avalanches in Ising systems are triggered by the flip of a single spin in a vanishing local field.   

We consider a given quenched realization of bonds $J_{ij}$, and analyze locally stable low-energy configurations of the system where each spin is aligned to the local field created by all other spins:
\begin{gather}
	\label{eqn:align}
	\vm_i = \frac{\vh_i}{|\vh_i|}.
\end{gather}
Here $\vm_i$ is the $T=0$ magnetization of spin $i$.
The local fields $\vh_i$ are defined as 
\begin{gather}
	\label{eqn:def-local-field}
	\vh_i = H\vex + \sum_j J_{ij}\vm_j.
\end{gather}
We are mostly interested in the dynamics at $T = 0$, where the magnetization $\vm_i$ within any local minimum becomes equal to the frozen spin direction: $\vm_i (T=0)= \vs_i$.  The Thouless-Anderson-Palmer equations obeyed by $\vm_i$ at finite temperature are discussed in Appendix~\ref{app:TAP}. 

\begin{figure}
	\includegraphics[width=0.9\columnwidth]{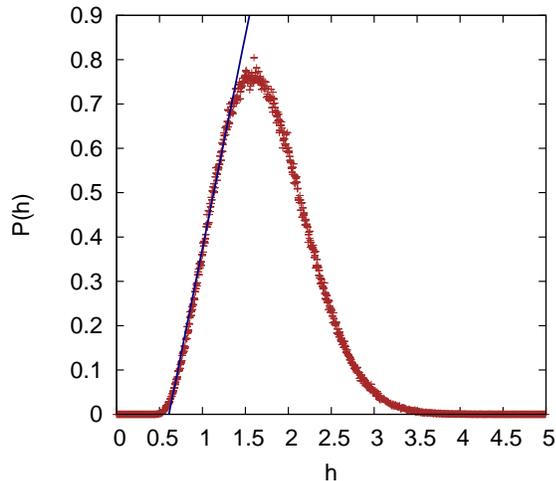}
	\caption{(Color online) Distribution $P(h)$ of the modulus of the local fields $h \equiv |\vh_i|$, obtained for a system size $N=1024$ and averaged over 1000 disorder realizations. The external magnetic field was set to $H=0$. A hard gap is clearly visible. The straight line is a fit ($f(h)= 0.96 (h-0.61)$) to the roughly linear increase of $P(h)$, as described in Ref.~\onlinecite{palmer1979internal}.}
	\label{fig:loc-fields-dist}
\end{figure}

The above relations, which are valid only at $T=0$, might appear to miss the contributions of the Onsager backreaction~\cite{thouless1977solution}, which, unlike in Ising systems at $T=0$, remains non-zero for vector spins at $T=0$~\cite{palmer1979internal}. Indeed, this finite Onsager term is known to be responsible for a hard gap in the distribution of local fields, as shown in Figure~\ref{fig:loc-fields-dist}, cf. Eq.~\eqref{eqn:def-local-field}. However, a careful analysis of the finite $T$ Thouless-Anderson-Palmer equations confirms that the analysis below does not miss any potential subtleties of the limit $T\to 0$. In particular, the inverse susceptibility matrix $A$ at $T=0$ is essentially identical to the one we obtain below in Eq.~\eqref{eqn:defA_mx} by working directly with the more naive equations given above (see App.~\ref{app:TAP} for details).

As the external field $H$ is increased, the magnetization increases smoothly by gradual readjustments of the spins, until a point of local instability is reached. At this point a larger discontinuous rearrangement is triggered upon further infinitesimal increase of $H$. We will describe the detailed dynamical rules applied in the event of a local instability in Sec.~\ref{sec:results}. Note that at $T=0$, in contrast to XY-spins, Ising systems do not display any adiabatic response, but only respond discontinuously by  magnetization avalanches, whenever an instability is triggered by a spontaneous spin-flip~\cite{pazmandi1999self}.

As we will see, the avalanche-like events triggered by local instabilities span a wide range of sizes. In fact the long-range character of the interactions $J_{ij}$ often induce system-spanning avalanches that involve a finite fraction of all spins. This contrasts with systems with short-range interactions for which it has been shown that single spin-flip dynamics (in Ising systems) does not lead to arbitrarily large, scale-free avalanches~\cite{andresen2013self}.

\subsection{Susceptibility and local instabilities}

Our main goal is to study the statistical properties of instabilities and avalanches generated in the evolution of the XY-glass, as it is progressively polarized. In Ising systems such instabilities are very easily identified by the criterion that a local field $h_i$ needs to vanish. A further infinitesimal increase of $H$ will then induce the corresponding spin to flip, potentially triggering an avalanche. For XY-spins local fields no longer easily identify instabilities, as they remain bounded away from zero~\cite{palmer1979internal}. Instead, one should study the susceptibility of the system to small changes in the external field $H$. Avalanche-like jumps will occur in configurations, in which the susceptibility diverges. This is equivalent to the inverse of the susceptibility matrix acquiring a zero eigenvalue, indicating that the system becomes soft.

Below, we derive the susceptibility and determine the condition for an instability, and thus a jump to occur in XY-spins. We define the local susceptibility to the external field $H$ as
\begin{gather*}
	\vchi_i = \frac{\partial\vm_i}{\partial H}.
\end{gather*}
By simple differentiation of Eq.~\eqref{eqn:align}, and using the definition of the local fields~\eqref{eqn:def-local-field}, we find the relation
\begin{gather}
	\label{eqn:basic}
	\chi_{i\mu} = \frac{1}{|\vh_i|}P_{\mu\nu}^i\left(\hat{e}_{x\nu} + \sum_j J_{ij}\chi_{j\nu}\right),
\end{gather}
where $P_{\mu\nu}^i = (\delta_{\mu\nu} - m_{i\mu}m_{i\nu})$ projects onto the direction orthogonal to the magnetization vector $\vm_i$. Summation of repeated indices $\nu$ is implied. In the above, Latin indices such as $i,j$ refer to sites, while Greek indices, $\mu,\nu \in \{x,y\}$, refer to spin components. Since at $T=0$ we have $\vm_i^2=1$, the magnetic response is always perpendicular to the instantaneous magnetization:
\begin{gather}
	\label{eqn:chi_perp}
	\vchi_i\cdot\vm_i = 0,
\end{gather}
as ensured by the projector in~\eqref{eqn:basic}. 

We can rewrite Eq.~\eqref{eqn:basic} using the property~\eqref{eqn:chi_perp} as:
\begin{gather}
	\label{eqn:mat_sec_model}
	\chi_{i\mu} = P_{\mu\nu}^i\frac{\hat{e}_{x\nu} + \sum_j J_{ij}P_{\nu\sigma}^j\chi_{j\sigma}}{|\vh_i|},
\end{gather}
which can be transformed into a matrix equation for the susceptibility $\vchi_i$:
\begin{gather}
	\label{eqn:defA_sec_model}
	\sum_{j\sigma} A_{i\mu,j\sigma} \chi_{j\sigma} = C_{i\mu},
\end{gather}
where
\begin{gather}	
	\label{eqn:defA_mx}
	A_{i\mu,j\sigma} = |\vh_i|\delta_{ij}\delta_{\mu\sigma} - J_{ij}\sum_\nu P_{\mu\nu}^i P_{\nu\sigma}^j,\\
	\label{eqn:defC_sec_model}
	C_{i\mu} = P_{\mu\nu}^i \hat{e}_{x\nu} = \delta_{\mu x}-m_{i\mu}m_{ix}.
\end{gather}
Note that the matrix $A$ is symmetric. In fact, as we confirm in App.~\ref{app:TAP}, it is the second derivative of the Gibbs free energy $G(\vm_i)$ with respect to $\vm_{i}$ and $\vm_{j}$, that is, the inverse of the susceptibility matrix in the $T\to 0$ limit.

With the help of the matrix $A$ we can formulate a criterion for local instabilities: The susceptibility $\vchi$ should diverge, i.e., the matrix $A$ should become degenerate and acquire a zero mode. 

The eigenvalues of $A$ corresponding to longitudinal eigenvectors, parallel to onsite magnetizations, can be computed analytically. From the definition of $A$, Eq.~\eqref{eqn:defA_mx}, it is immediate to check that the vectors
\begin{gather}
	\label{eqn:eig1}
	\vec{v}_i^{(k)} = \delta_{ki}\vm_i
\end{gather}
are eigenvectors of $A$ with eigenvalues $|\vh_k|$. As discussed above, the $|\vh_k|$ are always bounded from below by the positive Onsager term~\cite{palmer1979internal}, and hence there are no soft modes in the longitudinal sector of the spectrum. A more detailed discussion of the Onsager term at finite temperatures can be found in App.~\ref{app:TAP}. There the Onsager term reduces the "instantaneous fields" $|\vh_k|$ to the "thermodynamic fields" with equal orientation, but modulus $|\vy_k|=|\vh_k| - 1/(2h_\text{HM})$ where $h_\text{HM}^{-1}=2/N\sum_i 1/|\vy_i|$. The moduli of the fields $\vy_k$ are not bounded away from zero. However, the Hessian, its eigenvectors and eigenvalues at $T=0$ are not changed with respect to those obtained via the naive derivation at $T=0$.

The relevant soft modes are contained in the other half of the spectrum which corresponds to transverse response in the subspace orthogonal to the span of $\vec{v}^{(k)}$, $k=1,\dots ,N$. For every site $i$ we define the unit vector 
\begin{gather}
	\label{eqn:vecn}
	\vn_i\equiv\vm_i\times\vec{e}_z = (m_{iy},-m_{ix}), 
\end{gather}
which is  orthogonal to $\vm_i$. Since $\vchi_i$ is perpendicular to $\vm_i$ (see Eq.~\eqref{eqn:chi_perp}), we have $\vchi_i = \xi_i\vec{n}_i$ with $\xi_i = \vchi_i\cdot\vec{n}_i$. Projecting Eq.~\eqref{eqn:mat_sec_model} with $\vec{n}_i$ we obtain an equation in terms of $\xi_i$:
\begin{gather*}
	|\vh_i|\,\xi_i = m_{iy} + \sum_j J_{ij}\,\xi_j\,\vm_i\cdot\vm_j,
\end{gather*}
or equivalently,
\bea
	&&	\sum_j T_{ij}\,\xi_j = K_i,
\eea
where
\bea
	\label{eqn:defT_mx}
	&&	T_{ij} = |\vh_i|\delta_{ij} - J_{ij}\,\vm_i\cdot\vm_j,\\
	&&	K_i  = m_{iy}\notag.
\eea
Inverting, one finds the transverse susceptibilities, $\xi_j = T_{ji}^{-1}K_i$. 

The matrix $T$ is the inverse of the transverse susceptibility matrix. It is the central object in our study of instabilities and avalanches. The susceptibility diverges and the considered metastable state becomes locally unstable when $T$ acquires a zero mode. Below we denote by $e_k$ and $a_j^{(k)}$, for $k=1,...,N$  the eigenvalues and eigenvectors of $T$.

As one should expect, the requirement of spin alignment in a locally stable state, Eq.~\eqref{eqn:align}, is equivalent to imposing a local minimum (or saddle point) of the energy function $\mch$ (cf. Eq.~\eqref{eqn:ham_phi}) with respect to the angles $\{\phi_i\}$. The matrix $T_{ij}$ is in fact simply the Hessian of the Hamiltonian \eqref{eqn:ham_phi} with respect to the angles $\phi_i$,
\begin{gather}
	\label{eqn:HessianT}
	T_{ij} = \frac{\partial^2\mch}{\partial\phi_i\partial\phi_j},
\end{gather}
as we derive in more detail in App.~\ref{Tmatrix:deriv}. Stability requires the Hessian $T$ to be positive definite. An avalanche is triggered when its lowest eigenvalue becomes soft, $e_1\searrow 0$. 

\section{Numerical Analysis}
\label{sec:results}

\subsection{Dynamical protocol}

We have performed numerical simulations of the $T=0$ dynamics of the fully connected XY-spin-glass, Eq.~\eqref{eqn:ham}, with couplings drawn from the Gaussian distribution~\eqref{eqn:J_dist}. We have adopted simple dynamical rules that continuously decrease the energy until the system settles into a local minimum satisfying Eqs.~\eqref{eqn:align}. For a fixed value of the external field $H$, the $N$ spins are sequentially aligned with their local fields, as computed from Eq.~\eqref{eqn:align}, the local fields being updated on all other sites according to Eq.~\eqref{eqn:def-local-field} after each alignment. This procedure is iterated until the system converges to a local minimum of energy. Convergence is assumed if the state of the system does not change anymore (within numerical precision), and Eq.~\eqref{eqn:align} is satisfied for all spins. The external field is then increased by a small increment, and the above dynamics is repeated. 

To identify  the instabilities and the ensuing jumps along the hysteresis curve numerically, we monitor the lowest eigenvalue, $e_1$, of the inverse susceptibility matrix $T$. When $e_1$ reaches zero within numerical accuracy, we still need to ascertain that we deal with a genuine instability, and not some artifact due to numerical inaccuracy or insufficient convergence to the local energy minimum. To this end we drive the system back and forward by two increments of $H$, and determine whether the spin configuration changes strongly; if so, the event is accepted as a genuine jump.~\footnote{Sometimes $e_1$ becomes slightly negative, as a consequence of a not fully converged alignment procedure. In such cases, we apply a tiny noise and rerun the convergence algorithm to ensure that $e_1$ is eventually positive.}

We start the hysteresis loop at a large negative value $H = -H_0$ of the external field, such that the system is polarized and all the spins are aligned along the negative $x$-axis. In practice we chose $H_0 = 5$. The field $H$ is then gradually increased up to the large positive value $H = H_0$ where all spins point along the positive $x$-axis. To reproduce the adiabatic evolution as faithfully as possible we have used small increments of the field $\delta H = 0.005$ (independent of system size). This allowed us to  find all local instabilities, with the potential exception of very small jumps that are difficult to detect with the above described procedure. 

\subsection{Coercive field}

There is a critical magnitude of the field, $H_c$, at which the completely polarized state first becomes unstable and the magnetization departs from its extremal plateau. $H_c$ can be obtained by inserting the fully polarized state into the Hessian matrix in Eq.~\eqref{eqn:defT_mx} and determining the value of $H=H_c$ at which its lowest eigenvalue $e_1$ vanishes. More precisely, $H_c$ is the solution of the following equation:
\begin{gather}
	\label{eqn:def-Hc}
	\min\,\left(\textrm{spec}\left[\left |H_c + \sum_k J_{ik}\right|\delta_{ij} - J_{ij}\right]\right)=0.
\end{gather}
We note, however, that the value of $H_c$ depends on $N$ and diverges logarithmically in the thermodynamic limit.~\footnote{We conjecture that for finite size samples $H_c$ corresponds to the zero-temperature limit of the Gabay-Toulouse line~\cite{gabay1981coexistence}, which tends to $H_\textrm{GT}\to\infty$ as $T\to 0$.} 

At $H_c$ a transverse magnetization emerges, which spontaneously breaks the symmetry $y\leftrightarrow -y$. The corresponding rearrangement of magnetization is continuous, in contrast to the avalanches triggered by subsequent instabilities, which we will discuss below. Indeed one easily checks that the expansion of the energy $\mch(\{\phi\})$ around the fully polarized solution starts with a quadratic term, followed by quartic terms in the angular deviations from $\phi_i = \pi$. Thus the onset of transverse magnetization is qualitatively similar as the spontaneous symmetry breaking in a continuous mean-field phase transition, as described by Ginzburg-Landau theory.

The upward and downward branches of the hysteresis curve are found to coincide in the immediate vicinity of the fully polarized magnetization plateau. As shown in Fig.~\ref{fig:jumps_single} the upper plateau is seen to be reached at $H=+H_c$, precisely at the point at which the downward branch will start to deviate from the plateau. This coincidence is in contrast to the phenomenology in most Ising ferromagnets, where the extremal plateau is usually reached by a discrete final magnetization jump~\cite{jiles1986theory}.


\subsection{Avalanches}

\begin{figure}
	\includegraphics[width=\columnwidth]{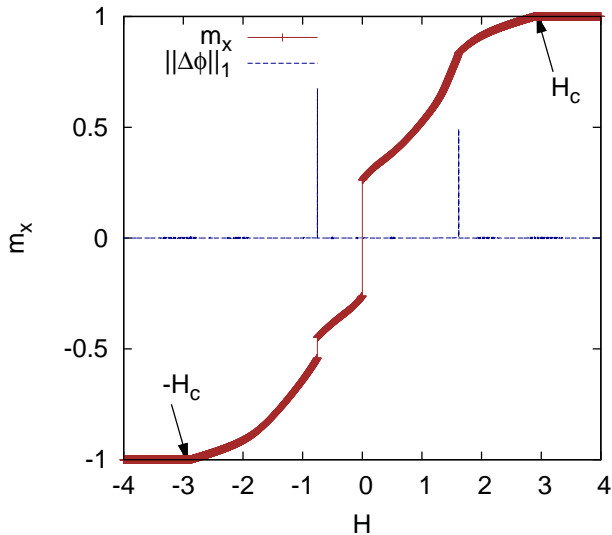}
	\caption{(Color online) The forward branch of the hysteresis loop in the average magnetization per spin, $m_x$, for a representative small sample of size $N=16$~(red, thick curve). At the points $H=\pm H_c$ the hysteresis curve starts deviating from full polarization. A big jump at $H=0$ arises since the magnetization spontaneously swivels by 180 degrees to realign with the external field which changes sign. Apart from that trivial jump,  two avalanche events are seen. Their magnitude $||\Delta\phi||_1$  is indicated by two peaks~(dashed, blue).}
	\label{fig:jumps_single}
\end{figure}

\begin{figure}
	\resizebox{0.95\columnwidth}{!}{\input{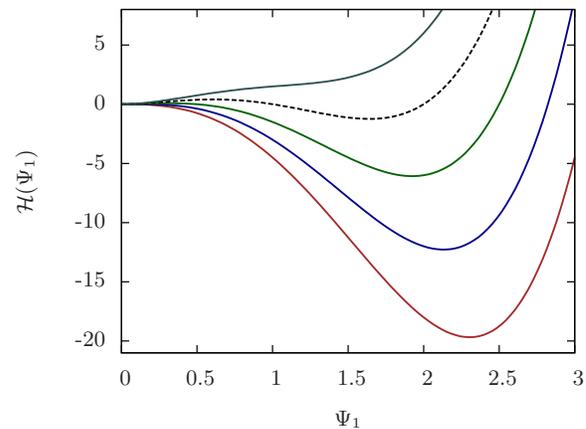}}
	\caption{(Color online) Schematic plot of the functional $\mch(\Psi=(\Psi_1,0,....,0)))$, Eq.~\eqref{eqn:V_dphi},  vs. $\Psi_1$, for various values of the external field $H$. The field increases from the top to the bottom curve. Before the instability ($H<H_\alpha$), $\mch(\Psi_1)$ has a locally stable minimum at $\Psi_1 = 0$. However, in general, there already exist lower lying minima at non-zero values of $\Psi_1$. As the field $H$ approaches the critical value $H_\alpha$ (the dashed curve), the minimum $\Psi_1 = 0$ becomes locally unstable and a spontaneous rearrangement (avalanche) is triggered.}
	\label{fig:lg-12}
\end{figure}

As illustrated in Fig.~\ref{fig:jumps_single}, between $-H_c$ and $H_c$, the polarization process consists in a succession of smooth sections of adiabatic magnetization, and avalanches that are triggered when a local instability occurs. These instabilities are very similar to spinodal lines at first order transitions. Indeed, let us expand the angular deviations $\delta\phi_i$ from a metastable state $\{\phi_i\}$ into the eigenmodes $a_i^{(k)}$ of the inverse susceptibility matrix $T$, $\delta\phi_i = \sum_k \Psi_k a_i^{(k)}$. The expansion of the energy around the local minimum then takes a Ginzburg-Landau form,
\begin{gather}
	\label{eqn:V_dphi}
	\mch(\{\Psi_k\}) = \sum_{k} A_k\Psi_k^2 + \sum_{klm} B_{klm}\Psi_k\Psi_l\Psi_m + \dots\,.
\end{gather}
The presence of the cubic term in the energy functional is a characteristic feature of first order transitions. It is responsible for a non-linear avalanche event, i.e., a discontinuous jump in $\Psi_k$, once the local minimum at $\Psi_k = 0$ becomes unstable, as illustrated in Fig.~\ref{fig:lg-12}. Note that generically the cubic term is non-zero. Only at $H=\pm H_c$ it vanishes, due to the symmetry of the polarized state. 

The instabilities which appear during the evolution of the external field induce avalanche-like rearrangements of the angles $\phi_i$. They manifest themselves  in the abrupt mesoscopic magnetization jumps seen in Fig.~\ref{fig:jumps_single}, where we show the upward branch of the full hysteresis loop of a small sample of size $N = 16$. This small size was chosen in order to display the essential avalanche features clearly. In the small sample one sees just two discontinuous magnetization jumps. Their size is measured by the average modulus of the change in the angle of the spins, 
\begin{gather}
	\label{eqn:jump_phi}
	||\Delta\phi||_1 = \frac{1}{N}\sum_i|\Delta\phi_i|.
\end{gather}
This avalanche characteristic is indicated by the peaks in Fig.~\ref{fig:jumps_single}. 
Note that $||\Delta\phi||_1$ can be rather large, even if in the course of the avalanche the magnetization  increases only by little, as is the case in the second avalanche of Fig.~\ref{fig:jumps_single}. This can happen when negative and positive changes in $\phi_i$ contribute nearly equally, such that the change in $m_x$ is small. 

\begin{figure}
	\includegraphics[width=0.95\columnwidth]{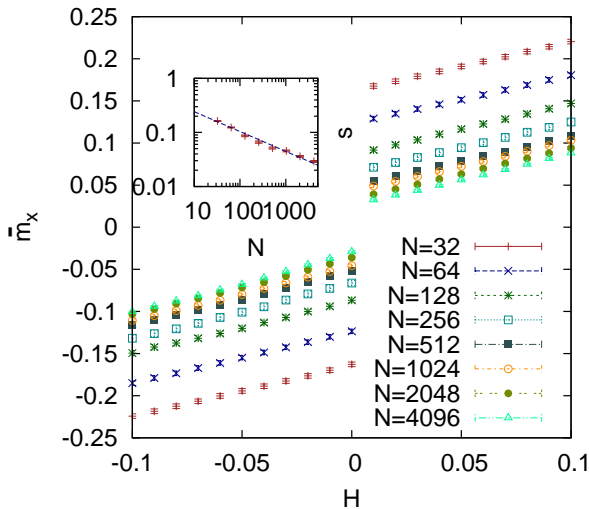}
	\caption{(Color online) The upward hysteresis curve, averaged over disorder for different system sizes $N = 32,64,128,256,512,1024,2048,4096$ (top to bottom curves for $H>0$ and bottom to top curves for $H<0$ respectively) with the number of disorder samples averaged over respectively being $1000,500,200,100,100,100,100,100$. We show the vicinity of $H=0$. The vertical span of the curve decreases with increasing $N$. \emph{Inset}: The average magnitude of the magnetization at $H=0^-$, $s \equiv |\overline{m_x}(0^-)|$ is a measure for the vertical span of the hysteresis curve. The decrease of $s$ (red points) with $N$ is consistent with a power-law $f(N) = b N^{-c}$. A fit yields $b = 0.56\pm0.05$ and $c = 0.37\pm 0.02$ (blue dashed curve in the inset).}
	\label{fig:jumpatorigin}
\end{figure}

\subsection{Jump at $H=0$ and subextensive width of the hysteresis loop}

At $H=0$, one always observes a large jump in magnetization. This has a trivial origin: at $H=0$ the energy $\mch$ is invariant under global rotations. If $m_{x}(H=0^-)<0$, an infinitesimal increase of $H\to 0^+$ will induce the entire magnetization pattern to swivel around by $180$ degrees and align with the positive field. The magnitude of the zero field magnetization, $s\equiv |m_{x}(H=0^-)|$, is a measure of the vertical span of the hysteresis curve. This span is a measure of how strongly off-equilibrium the system is driven. While in usual ferromagnets the span is finite in the thermodynamic limit, i.e., the magnetization differs extensively from its equilibrium value, we find here that the span scales to zero with increasing system size $N$. Figure~\ref{fig:jumpatorigin} shows the sample-averaged magnetization per spin on the hysteresis curve for various system sizes. The decrease of $s$ fits well to a power law decay $f(N) = b N^{-c}$ with $c\approx 0.37\pm 0.02$, as shown in the inset of Fig.~\ref{fig:jumpatorigin}. This behavior is very similar to the weak empiric power law decay $N^{-x}$ of the width of the hysteresis loop in the Ising spin-glass, where we found an exponent $x\approx0.2$ from fitting simulation data. However, we note that the data are also compatible with logarithmic scaling. The fact that in both the XY- and Ising glass the hysteresis loop has no extensive width in fully connected models seems not to have been noticed in previous studies. It indicates that the quasi-adiabatic dynamics is probing states that are in fact still comparatively close to equilibrium\cite{aspelmeier2004complexity}.

It is interesting to note that this  phenomenon is quite similar to what has been predicted analytically for the long-time Langevin dynamics in the SK model at finite temperature~\cite{cugliandolo1994on,cugliandolo1995weak}, and is observed numerically in simulations in fully connected spin-glasses: While the glassy system is definitely out-of-equilibrium and undergoes slow aging dynamics in phase space, the energy density and any other extensive thermodynamic observables approach their equilibrium values very closely, up to sub-extensive corrections. This happens even quite rapidly following an initial relaxation. Here we find a close analogue of this behavior at strictly zero temperature, under adiabatically slow driving.

\subsection{Avalanche observables}

In Ising systems, magnetization avalanches are almost completely characterized by two quantities: the increase  of the total magnetization, $\Delta M$, and the size of the avalanche, $S$, that is, the number of  spins that flip during this avalanche~\cite{pazmandi1999self,andresen2013self,vives1995universality}. In contrast, glasses with continuous symmetry are richer in the sense that they allow for a finer characterization of the avalanches and their relation with the inverse susceptibility matrix just before the avalanche is triggered. 

\subsubsection{Magnetization jump and avalanche size}

Apart from the change in the $x$-component of the total magnetization $\Delta M_x$, 
\begin{align}
	\label{eqn:delMx}
	\Delta M_x &= \sum_i\Delta m_{ix} = N\Delta m_x,
\end{align}
we also monitor the magnitude of the change in the magnetization vector $|\Delta\vec{M}|$, 
\begin{align}
	\label{eqn:delvecM}
	|\Delta\vec{M}| &= \left|\sum_i\Delta\vm_i \right|.
\end{align}
The notation $\Delta X$ denotes the difference of the quantity $X$ in the metastable configuration just after and before the avalanche.

\begin{figure}
	\includegraphics[width=0.95\columnwidth]{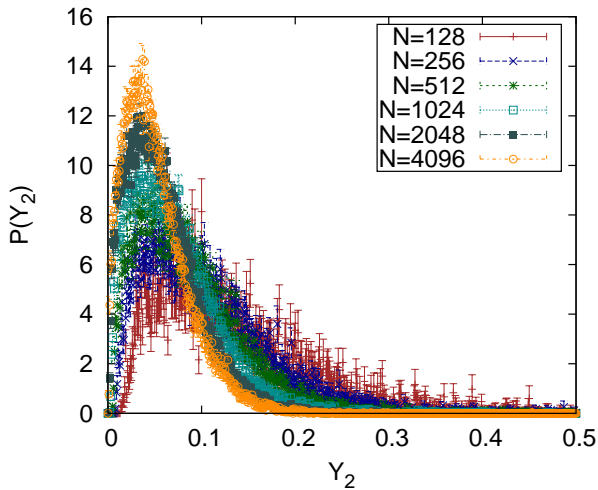}
	\caption{(Color online) The distribution of the participation ratio $Y_2$, as defined by Eq.~\eqref{eqn:jump_PR} for several system sizes $N$. The peak of the distribution increases with $N$ and shifts towards $Y_2=0$. Even though the thermodynamic limit is not yet clearly reached at $N=4096$ (curve with the highest peak [orange]), the data suggests that, as $N\to \infty$, $P(Y_2)$ remains peaked at a finite $Y_2\approx 0.05$. This would imply that  there is a typical avalanche size of order $N$.}
	\label{fig:jumps_PR}
\end{figure}

In order to characterize the fraction of spins effectively involved in an avalanche, we consider the participation ratio $Y_2$, defined as:
\begin{align}
	\label{eqn:jump_PR}
	Y_2 & = \frac{1}{N}\frac{\left[\sum_i(\Delta m_{ix})^{2}\right]^{2}}{\sum_i(\Delta m_{ix})^{4}}.
\end{align}
We evaluated its probability distribution $P(Y_2)$ over all avalanches in a given sample. The sample-averaged $P(Y_2)$ is plotted in Fig.~\ref{fig:jumps_PR}. The data suggests that a finite fraction of order $\sim 0.05$ of all spins participates in a typical jump, while avalanches that are much smaller than the system size are rare. Interestingly this differs from avalanches in the Ising SK model, where the density of small avalanches diverges as an inverse power law of the avalanche size. The latter can be seen as a form of self-organized criticality of those Ising systems.

\subsubsection{Fraction of avalanches in the magnetization process}

\begin{figure}
	\includegraphics[width=0.95\columnwidth]{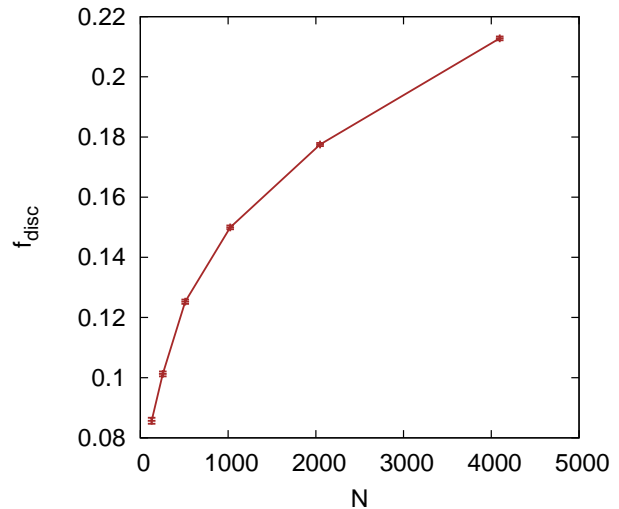}
	\caption{(Color online) Fraction of the total magnetization reversal, which occurs in the form of discontinuous avalanches. In the thermodynamic limit, the curve saturates, conceivably to a fraction less than $1$. However, larger system-sizes would be needed to yield a reliable estimate of $f_\textrm{disc}(N\to\infty)$.}
	\label{fig:jumps_fraction}
\end{figure}

Ising spins at $T=0$ can adjust to a change of external field only by discontinuous spin-flips and avalanches. In contrast, systems with continuous degrees of freedom continue to polarize under an increase of the external field, even between discontinuous jumps, as seen in Fig.~\ref{fig:jumps_single}. It is thus interesting to ask, what fraction of the polarization reversal along the upward hysteresis branch is due to discontinuous jumps and adiabatic polarization, respectively. For the XY-glass, we studied numerically the fraction $f_\textrm{disc}$ due to avalanches. Fig.~\ref{fig:jumps_fraction} shows that $f_\textrm{disc}$ increases with the system size, but presumably saturates to some finite value $f_\textrm{disc}(N\to\infty) < 1$, since the linear susceptibility between avalanches remains of order $O(1)$. The survival of a finite fraction due to continuous events was predicted in a different system, namely pinned elastic manifolds and their static, \emph{equilibrium} evolution under an external force~\cite{le2010avalanches,ledoussal2012equilibrium}. In order to determine the limiting fraction for the XY glass as $N\to\infty$, however, one would have to perform simulations of larger systems than we were able to study.


\subsection{Marginal stability: Gapless spectrum of the inverse susceptibility matrix}

\begin{figure}
	\includegraphics[width=0.95\columnwidth]{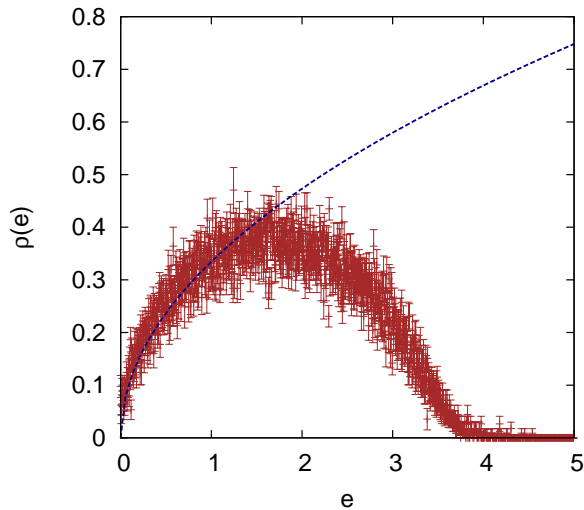}
\caption{(Color online) Average spectral density of the inverse susceptibility matrix $T$, averaged over all instability points occurring for $|H|<1$ (except the trivial jump at $H=0$) in a single disorder sample of size $N=1024$ (red dots). The blue dashed curve is a fit to the function $\rho(e) = a\sqrt{e}$ in the region $[0,1.5]$, with $a = 0.335\pm 0.003$. This confirms the gapless `semicircle law'~\eqref{eqn:semicircle} for small eigenvalues.}
	\label{fig:jumps_wigner}
\end{figure}

We have already discussed that the lowest eigenvalue of the inverse susceptibility matrix $T$, $e_1$, vanishes at an instability. It is also of interest to analyze the remainder of the spectrum of $T$ along the hysteresis curve. The spectral density of $T$, averaged over critical metastable states (just before an instability) is shown in Fig.~\ref{fig:jumps_wigner}. The distribution at small eigenvalues is given by the edge of a semicircle law~\cite{mehta2004random},
\begin{gather}
	\label{eqn:semicircle}
	\rho(\lambda)\sim\sqrt{\lambda}.
\end{gather}
This is reminiscent of the spectrum of Hessians found in the dynamics of fully connected glasses~\cite{cugliandolo1993analytical}. However, it is very different from the rather pathological spectra, which one finds for Hessians evaluated on metastable solutions of Thouless-Anderson-Palmer equations at extensive energies above the ground states~\cite{yeo2004complexity,muller2006marginal}. This is again consistent with the finding that our adiabatic spin alignment dynamics remains subextensively close to the ground state, and does not explore the regime of high excitation energies, which are presumably irrelevant for physical dynamics.

Between the jumps the inverse susceptibility matrix has a small positive gap $e_1>0$. 
However, the gap never becomes large, but rather scales as 
\begin{gather}
	\label{eqn:e1}
	e_1\sim N^{-2/3},
\end{gather}
being of the same order as the level spacing between $e_1$ and $e_2$, given the spectral density~\eqref{eqn:semicircle}. This is similar to what is found in the analysis of metastable states at $T=0$~\cite{yeo2004complexity}.
 The above may be seen as the analogue of the fact that in metastable states of the Ising spin-glass the smallest local field always remains of the order of $N^{-1/2}$, which is of the same order as the difference between the smallest two local fields. In this sense both glassy systems are thus \emph{marginally stable}, having a stability towards perturbations which vanishes in the thermodynamic limit. This feature is not unexpected, since,  at least at equilibrium, the continuously broken replica symmetry of the spin-glass phase ensures the presence of massless replicon modes and thus criticality.

\subsection{Density of avalanches and fractality of soft modes}

\begin{figure}
	\includegraphics[width=0.95\columnwidth]{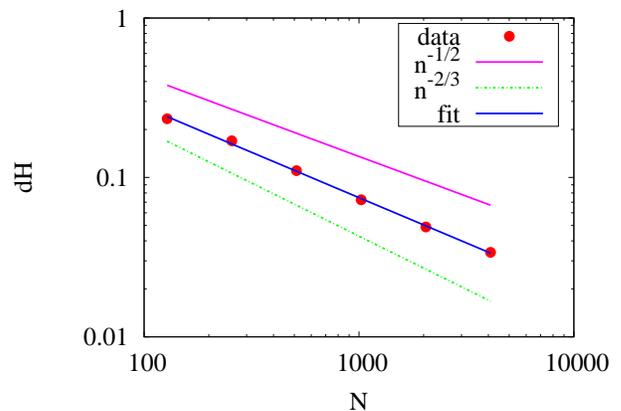}
	\caption{(Color online) The log-log plot of the average inter-avalanche spacing $\delta H$ within the range $|H|\leq 1$ of external fields. The (red) dots are the numerical data. For comparison we show the two power laws $N^{-1/2}$ (top line [black]) and $N^{-2/3}$ (lower line [brown]), that are suggested by scaling arguments. The best fit to the numerical data is $\delta H\sim N^{-a}$ with $a=0.57\pm 0.01$ (middle line [blue]).}
	\label{fig:jumps_per_H}
\end{figure}

Despite the different scalings and the different nature of the trigger of avalanches in XY- and Ising glasses, the discrete values $H_\alpha$ of the external field, at which avalanches occur, appear to be spaced by similar orders of magnitude, $\delta H_\alpha = H_{\alpha+1}-H_\alpha\sim N^{-\alpha}$ with $\alpha \approx 1/2$. For the Ising case, this was established numerically in Ref.~\onlinecite{pazmandi1999self}, and $\alpha=1/2$ was shown to be the exponent arising in equilibrium shocks in Refs.~\onlinecite{le2010avalanches,ledoussal2012equilibrium}. For the XY-glass the numerical data in Fig.~\ref{fig:jumps_per_H} shows that the number of avalanches per unit of the external field (for $0<|H|<1$) is consistent with a scaling $N^{\alpha}$ with $\alpha\approx 0.57$.

Theoretically, one might anticipate a scaling $\delta H_\alpha\sim N^{-1/2}$ based on the following heuristic consideration. In the SK glass it was found that the numerically studied out-of-equilibrium avalanches are distributed with the same power laws as static magnetization jumps in the ground state configuration, and both feature a typical number $O(N^{1/2})$ of avalanches/jumps  per unit increment of the field. This is presumably a consequence of the before-mentioned fact that the dynamics remains in a certain sense close to equilibrium. If the same similarity holds in the XY-glass, we may conjecture the scaling of $\delta H$ based on such static considerations. Those are in fact the same as in the Ising model: Replica symmetry breaking in the spin-glass phase suggests that there are a number low lying states with energy difference of $\Delta E =O(1)$, whose spin orientation is, however, macroscopically different (with an overlap strictly smaller than $1$). Their total magnetizations $M_x$ are expected to differ by subextensive fluctuations $\Delta M_x =O(\sqrt{N})$. From this one expects the ground state to jump as soon as the external field is varied by a quantity of order  
\begin{gather}
	\label{eqn:dHstatic}
	\delta H_{\rm static} =\Delta E/\Delta M_x \sim N^{-1/2}.
\end{gather}

However, in order to better understand these scalings in the \emph{dynamics} of the XY-glass, we estimate the typical distance between avalanche-like events, $\delta H_{\rm dyn}$, with simple scaling arguments. In the local minimum of the anlges $\{\phi_i\}$ which the system reaches via an avalanche right after an instability, the inverse susceptibility matrix $T=T_0$ is expected to have a lowest eigenvalue $0<e_1\sim N^{-2/3},$. The corresponding soft mode is likely to drive the next instability as we increase the field further by $\delta H$. Expanding the deviation from $\{\phi_i\}$ as $\delta \phi_i = \sum_k \Psi_k a_i^{(k)}$, where the $a_i^{(k)}$ are the eigenmodes of $T_0$, and expanding the energy $\mch$ as a function of the $\Psi_k$, we find
\begin{gather}
	\mch = \delta H \sum_k b_k\Psi_k + \sum_{jk}\frac{e_k\delta_{jk}+\delta H\, c_{jk}}{2} \Psi_j\Psi_k+ \dots\, ,
\end{gather}
where 
\bea
	b_k &=& \sum_i a^{(k)}_{i} \sin(\phi_i),\\
	c_{jk} &=& \sum_i a^{(j)}_{i}a^{(k)}_{i} \cos(\phi_i).
	\label{eqn:ckk}
\eea

\begin{figure}
	\includegraphics[width=0.95\columnwidth]{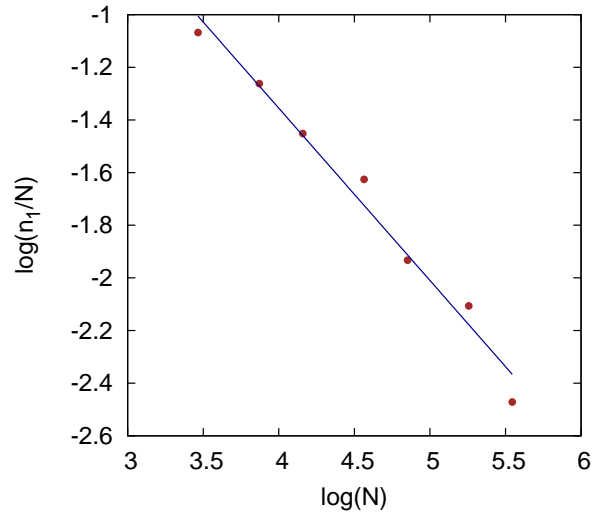}
	\caption{(Color online) Participation ratio of the soft mode divided by the system size, $n_1/N$, plotted versus $N$. The (red) dots are the numerical values, the solid (blue) curve is the fit to the data: $\ln(n_1/N)  = 1.26 - 0.65 \ln(N)$, which is compatible with the theoretically anticipated scaling $n_1\sim N^{1/3}$.}
	\label{fig:pr-n1}
\end{figure}

The next instability is expected when the first eigenvalue of the perturbed Hessian, $e_k\delta_{jk}+\delta H c_{jk}+O(\Psi)$, turns zero. To leading order in $\delta H$, the eigenvalues are simply shifted as $e_k'= e_k+\delta H\, c_{kk}$. Thus we expect, to leading order at large $N$, the distance between avalanches to be given by
\begin{gather}
	\delta H_\alpha = {\rm min}_{k, c_{kk} <0} \left[\frac{e_k}{c_{kk}}\right].
\end{gather}
As mentioned above, the smallest eigenvalues $e_k$ scale as $N^{-2/3}$. The coefficients $c_{kk}$  are more subtle to estimate, and their scaling may in fact depend on the location along the hysteresis loop. To estimate the sum in Eq.~\eqref{eqn:ckk}, we first need to analyze the structure of the softest eigenvectors. Interestingly, they are neither fully localized, nor completely delocalized. Instead they are fractals, having an inverse participation ratio, which we empirically find to scale as
\begin{gather}
	\label{eqn:pr-n1-N}
	\frac{1}{n_k} \equiv \sum_i [a_i^{(k)}]^4 \sim \frac{1}{N^{1/3}}.
\end{gather}
This is extracted from the numerical data in Fig.~\ref{fig:pr-n1}, where we show the average participation ratio $n_1$ of the softest eigenmode $e_1$ as a function of $N$ (averaged over the hysteresis loop in the range $0.01<|H|<1$). We checked that the scaling of higher moments, $\sum_i [a_i^{(k)}]^{2q}$ is consistent with $n_1^{1-q}$, that is, there are no indications of multi-fractality of those modes.  


The above suggests that we may think of the terms $[a_i^{(k)}]^{2}$ in~\eqref{eqn:ckk} as being of order $1/n_1$ on $O(n_1)$ sites, while being negligible in the bulk of the system. For avalanches in the low field region, where $|H|\ll 1$ and $\overline{\cos(\phi_i)} = m_x\ll 1$, we further assume that  on the relevant $O(n_1)$ sites  the magnetization $m_{ix}=\cos(\phi_i)$ is randomly signed. From this we finally expect the scaling  
\begin{gather}
	\label{eqn:ckk2}
	c_{kk}\sim n_k^{-1/2} \sim N^{-1/6}.
\end{gather}
Together with~\eqref{eqn:e1} this then suggests the scaling
\begin{gather}
	\label{eqn:deltaHscaling}
	\delta H_\textrm{dyn}\sim\frac{e_k}{c_{kk}} \sim N^{-2/3} n_k^{1/2}\sim N^{-1/2},
\end{gather}
at least at small $H$. This is indeed in agreement with the expectation~\eqref{eqn:dHstatic} from static considerations. At larger $H$, however, where the magnetization is extensive, it is not clear that $\cos(\phi_i)$ on the relevant sites for the softest mode can be considered random in sign. One might then rather expect $c_{kk}\sim O(1)$ and thus a trend to see $\delta H_\textrm{dyn}\sim N^{-2/3}$.
  
A numerical study of the scaling of the avalanche-averaged coefficient $c_{11}$ with $N$ was too inconclusive to allow us to establish the scaling~\eqref{eqn:ckk2} directly. A possible reason is that the scaling indeed depends on the proximity to zero magnetization, in which case the averaging over avalanches in a finite window of $H$ would result in inconclusive scalings with $N$. These considerations might also be the reason why the total number of avalanches within $H\in [-1,1]$ was found to scale like $N^\alpha$, cf. Fig.~\ref{fig:jumps_per_H}, with the best fitting exponent $\alpha=0.57\pm0.01$ being intermediate between the scalings one may expect close to $H=0$ and at finite $H$.
 
To conclude this discussion, it is interesting to note that, if one \emph{assumes} the scalings in Eqs.~(\ref{eqn:dHstatic},\ref{eqn:e1}) as given, as well as the scaling $c_{kk}\sim n_k^{-1/2}$, one could \emph{predict} the fractality~\eqref{eqn:pr-n1-N} of the soft modes, $n_k\sim N^{1/3}$. Obviously, it would be interesting to derive this fractality directly, without invoking the various heuristic arguments above. 

We point out that the fractality of the softest eigenvector of the spin-glass Hessian is not a trivial finding. If one considers the Hessian~\eqref{eqn:defT_mx} as essentially a random Gaussian matrix, apart from some shifts on the diagonal, one might expect the eigenvectors to behave like in the standard Gaussian matrix ensembles, namely as $n_k\sim N$. The fact that this is not true implies that the Hessians of spin-glass minima are distinctly different from standard random matrix ensembles. 

\subsection{Statistics of avalanches}

A numerical study of hysteresis in the fully connected Ising spin-glass~\cite{pazmandi1999self} displayed self-organized criticality throughout the hysteresis loop, the distribution of avalanche sizes being a scale-free power law, cut off only by the system size. Self-organized criticality~\cite{bak1987self,bak1988self} is said to occur in a system if, without fine-tuning, it acquires critical behavior, such as widely distributed, scale-free response, as a consequence of the dynamical evolution towards a critical attractor. 

The criticality of the fully connected SK model contrasts, however, with short-ranged Ising systems, such as the random field Ising model~\cite{sethna1993hysteresis,perkovic1995avalanches,vives1995universality} or Edwards-Anderson spin-glasses in finite dimensions~\cite{andresen2013self}, which display criticality only upon fine-tuning the strength of disorder and the value of the external field. In the SK model, criticality arises due to the long (infinite) range of interactions: The flip of a single spin has a finite probability of inducing other spin-flips and thereby triggering a large avalanche. Since in the SK model the spin-flips are not confined to a small neighborhood of the original spin, the avalanche may spread up to sizes which diverge with the system size. 


It is interesting to see whether this intriguing criticality and system spanning avalanches are also present in systems with continuous degrees of freedom, as considered here. Naturally, it is to be expected that long-range interactions are again crucial, as is also suggested by studies on random field XY-models~\cite{silveira1999critical}. As we will show below the avalanches are still system spanning, but they are \emph{typically} of the system size, and do not display a scale-free power law which decreases with increasing system size. 

We have analyzed the statistics of several avalanche characteristics, such as the magnetization jump $\Delta M_x$ and the size of the avalanche, as measured by $|\Delta\vec{M}|$. More precisely, we have calculated the frequency of occurrence of a given avalanche observable, let us call it $X$, per unit of external field and unit interval in $X$, upon averaging over disorder,
\begin{gather*}
	\rho(X)\equiv\frac{1}{\Delta H \,\delta X} \sum_{H_\alpha\in [H-\Delta H/2 ,H+\Delta H/2]} \overline{\chi_{[X,X+\delta X]}(X_\alpha) }.\\
\end{gather*}
In this formula $H_\alpha$ are the values of external fields at which instabilities occur, and $X_\alpha$ are the associated avalanche observables. $\chi_{[a,b]}$ denotes the characteristic function of the interval $[a,b]$, and the overbar denotes the disorder average. 
Note that in typical samples the sum is expected to contain a number of terms of the order of $N^{\alpha}\Delta H \delta X$, with $\alpha \approx 1/2$, as discussed above. Therefore we expect that $N^{-\alpha} \rho(X)$ has a proper thermodynamic limit, upon which one may shrink the increment $\delta X\to 0$. One could also take the limit $\Delta H\to 0$, and study $\rho(X)$ as a function of the external field $H$. However, here we content ourselves with an analysis of the avalanche statistics in a finite interval, setting $H=0$ and $\Delta H=2$, but excluding the huge jump at $H=0$ which we discussed previously. Notice that we do not normalize these densities, that is, we do not impose $\int dX \rho(X)= 1$, otherwise we would loose information about the frequency with which avalanches occur as $H$ increases. 

For system sizes $N \leq 2048$, we generated $1000$ samples of disorder, while for the largest systems, $N = 4096$, we considered $714$ samples. 

\begin{figure}
	\includegraphics[width=0.95\columnwidth]{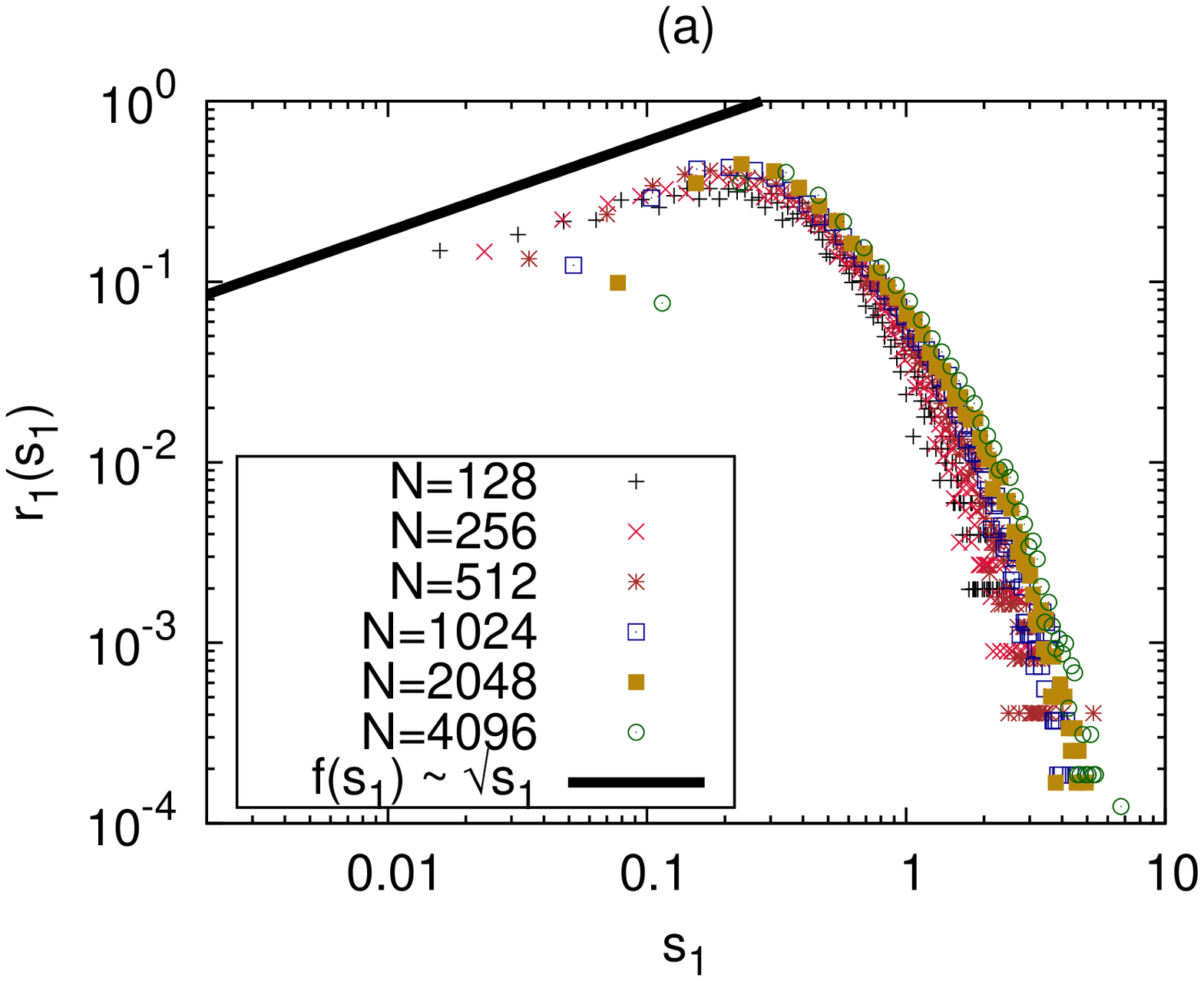}
	\includegraphics[width=0.95\columnwidth]{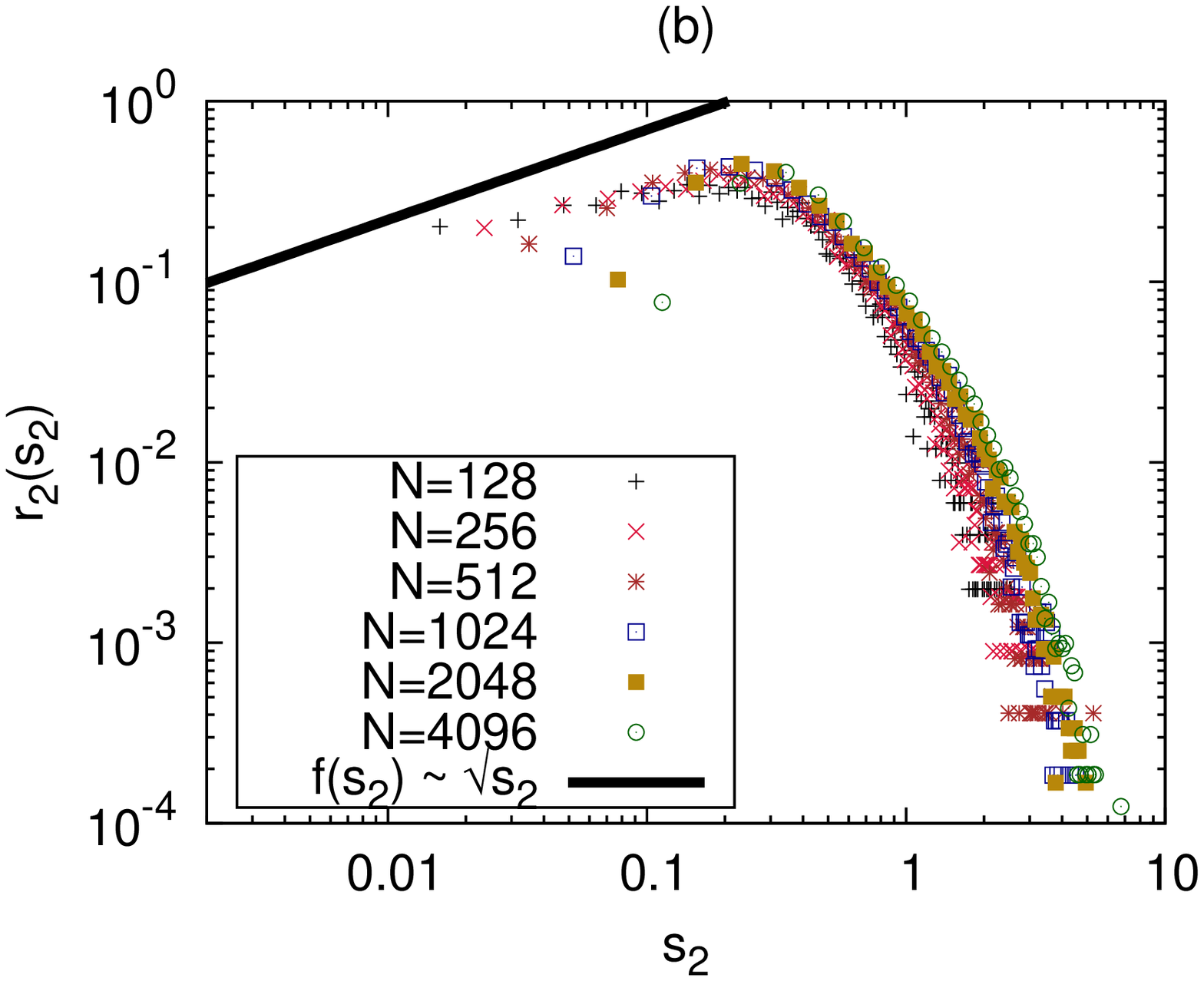}
	\caption{(Color online) Data collapse of scaled densities of two different measures of avalanches: (a) $s_1 = N^{-a}\Delta M_x$, $r_{1}(s_{1}) = N^{2a-1}\rho(\Delta M_x)$; (b) $s_{2} = N^{-a}|\Delta \vec M|$, $r_{2}(s_{2}) = N^{2a-1}\rho(|\Delta\vec{M}|)$. The body of the data collapses best with an exponent $a=1-\alpha=0.43$, which is in agreement with $\alpha=0.57$ obtained in Fig.~\ref{fig:jumps_per_H}. The solid (black) line corresponds to a power law $\sim s^{1/2}$.}
	\label{fig:jump_densities}
\end{figure}

We assume that in a finite size system the distribution of the observable $X$ has a cut-off which scales as $N^a$, where $a$  depends in general on the observable. It is then natural to define the rescaled variable $s \equiv N^{-a} X$. As argued above, we expect $O(N^{\alpha})$ avalanches per unit interval of $H$. Thus we define the rescaled density $r(s)$:
\bea
	\label{eqn:rs}
	r(s) &\equiv &   \frac{N^{-\alpha}}{\Delta H \delta s} \sum_{|H_\alpha|< \Delta H/2} \overline{\chi_{[s,s+\delta s]}(s_\alpha \equiv N^{-a}X_\alpha) }\notag\\
	&=&   \frac{N^{-\alpha}}{\Delta H \delta s} \sum_{|H_\alpha|< \Delta H/2} \overline{\chi_{[N^as,N^as+N^a\delta s]}(N^{a}s_\alpha\equiv X_\alpha) }\notag\\
	&=&   \frac{N^{a-\alpha}}{\Delta H \delta X} \sum_{|H_\alpha|< \Delta H/2} \overline{\chi_{[N^as,N^as+\delta X]}(X_\alpha) }\notag\\
	&=&   N^{a-\alpha} \rho(X=N^a s),
\eea
which we expect to have a well-behaved limit as $N\to\infty$, provided the value of the exponent $a$ is chosen appropriately. Considering that a finite fraction of the magnetization process occur in avalanches, and that the bulk of the increase of $M_x$ from $-N$ to $N$ occurs over a range of order $1$ in $H$, through a number $N^\alpha$ of avalanches, it is natural to expect that the typical scale for $\Delta M_x$ is $N^{1-\alpha}$, i.e., $\alpha=1-a$. Essentially the same scaling appears to apply to $|\Delta \vec M|$ as well. 
Thus, we attempt a scaling plot of $N^{2a-1}X$ vs. $s= N^{-a} X$, treating $a$ as a free exponent to be fitted. For both observables $X=\Delta M_x$ and $X=|\Delta \vec M|$, we found the best data-collapse for the function $r(s)$ with the cut-off exponent $a= 1-\alpha= 0.43$, as shown in Fig.~\ref{fig:jump_densities}. This is in good agreement with the exponent $\alpha=0.57$ obtained from the scaling of the density of avalanches in Fig.~\ref{fig:jumps_per_H}.

The scaling plots in Fig.~\ref{fig:jump_densities} show that both quantities $X=\Delta M_x$, and $X=|\Delta\vec{M}|$ are reasonably well described by scaling laws. However, unlike analogous distributions in Ising glasses, their distribution does not display scale-free behavior with a decreasing power law. Rather, small avalanches are rare and the bulk weight of the distribution sits at the cut-off scale. Note also that for small values of the scaling variable $s$, the scaling collapse is rather poor. We attribute this to difficulties in the detection of those small jumps. As we described earlier, we had used the presence of local hysteresis as a necessary criterion to qualify a candidate avalanche as a genuine instability. However, this test is not rigorous for very small jumps with a magnetization change comparable to that of the typical smooth increase of magnetization over an interval of the length of our numerical increment $\delta H = 0.005$. Thus, the densities for small jumps $\Delta M_x = N^a s \lesssim  N \delta H$, i.e., for $s\lesssim 0.005\times N^{1-a}$, are not really reliable.   
 
We conjecture instead that the true densities should also scale at small $s$. In fact, a power-law  $r(s)\sim s^{\gamma}$ with $0<\gamma\approx 0.5$ seems to describe relatively well the data for smaller $N$, where we have higher confidence in our small-$s$ statistics. Such an increasing power-law is quite in contrast with the decreasing power law $\sim 1/s$ in the Ising glass, which implies a scale-free avalanche distribution in that system. On the other hand, a similar increasing power law (but with different exponent, $\sim s$) is found in the distribution of \emph{equilibrium} jumps of mean-field systems that display one-step replica symmetry breaking~\cite{le2010avalanches}.

Given that the low-$T$ limit of the Parisi solution for vector spin-glasses is not well understood to date, we may our out-of-equilibrium findings to make a conjecture about the nature of replica symmetry breaking in these glasses. Let us assume for a moment that the XY-glass is described by continuous replica symmetry breaking and an order parameter function with a low-$T$ limit behaving as $q(x\gg T)\approx 1-c (T/x)^\mu$ with $\mu>0$, similarly as in the Ising glass, where $\mu=2$. Then the analytical results of Ref.~\onlinecite{le2010avalanches}, generalized to the present case, predict equilibrium jump distributions with a \emph{decreasing} power law $\rho(\Delta M_x) \sim \Delta M_x^{-\tau}$ with exponent $\tau =2/\mu$. If one further stipulates that dynamic and static avalanches behave similarly in systems with continuous RSB, as it happens in the Ising case, this would be inconsistent with our numerical findings. This leads us to conjecture that the replica symmetry breaking at low temperature in the XY-glass is not simply continuous (sometimes referred to as "full replica symmetry breaking"). On the other hand, there is definitely such a continuous replica symmetry breaking at temperatures below but close to $T_c$, and it appears unlikely that it would turn into a simple one-step phase at lower $T$~\cite{gabay1981coexistence,cragg1982instabilities,gabay1982symmetry,sharma2010almeida}. A more likely scenario might be a low $T$ transition to a phase with a $1+$FRSB structure, where the overlap function $q(x)$ has a discontinuity at large $q$, as it was found in spin glasses with mixed spin interactions~\cite{crisanti2006spherical}.


\subsection{Role of the soft mode in the jumps}

Since jumps are triggered by a single mode which becomes soft at the instability, it is natural to ask how much the (non-linear) jump is actually correlated with the soft mode which triggers it. An analogous problem was investigated in the context of jamming~\cite{brito2007heterogeneous}, where a strong correlation between the few softest modes of the corresponding inverse susceptibility matrix and the ensuing avalanche was found. Here we find a very similar situation: in an avalanche the softest linear modes contribute most. Below we quantify this in more detail.

We define the $N$-dimensional vector of magnetization jumps $\vec{Z}=(\Delta\vm_1,\Delta\vm_2,\cdots,\Delta\vm_N)$. The two-dimensional $\Delta\vm_i$ and the jumps in the angles, $\Delta\phi_i$, are simply related by:
\begin{gather}
	\Delta\vm_i = -(1 - \cos\Delta\phi_i)\vm_i - \sin\Delta\phi_i \,\vn_i,
\end{gather}
where $\vn_i$ was defined in Eq.~\eqref{eqn:vecn}. As we discussed in Sec.~\ref{sec:model}, the spectrum of the inverse susceptibility matrix $A$ splits naturally into longitudinal eigenvectors, Eq.~\eqref{eqn:eig1}, and transverse eigenvectors, given by the spectrum of the Hessian \eqref{eqn:HessianT}. Denoting them $\vec{v}_{Lj}$ and $\vec{v}_{Tj}$ respectively, $j=1,...N$, we have the following decomposition of unity:
\begin{gather}
	\label{eqn:unity_decomposition}
	1 = \sum_{j=1}^{N} \frac{(\vec{Z} \cdot\vec{v}_{Lj})^2}{|\vec{Z}|^2} + \sum_{j=1}^{N} \frac{(\vec{Z} \cdot\vec{v}_{Tj})^2}{|\vec{Z}|^2} \equiv \sum_{j=1}^{N} \omega_{Lj}^2 + \sum_{j=1}^{N} \omega_{Tj}^2,
\end{gather}
where $\omega^2_{Lj} = (\vec{v}_{Lj} \cdot \vec{Z}/| \vec{Z}|)^2$ and $\omega_{Tj}^2 = (\vec{v}_{Tj} \cdot \vec{Z}/| \vec{Z}|)^2$ are the contributions due to longitudinal (L) and transverse (T) modes, respectively. 

\begin{figure}
	\includegraphics[width=0.95\columnwidth]{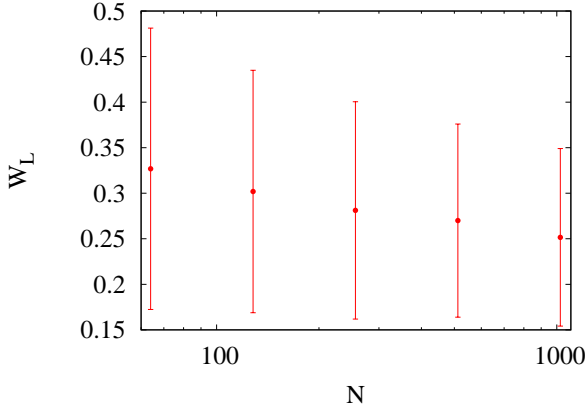}
	\caption{(Color online) Average contribution $W_L$ (see Eq.~\eqref{eqn:Fpar}) of longitudinal modes to the avalanches, plotted as a function of system size. Note the large standard deviations, indicating that jumps come in all sizes.}
	\label{fig:WL}
\end{figure}

We quantify the contribution of a set of linear modes to a magnetization jump $\vec{Z}$ by the total weight of that set in the decomposition. The total contribution from longitudinal modes can be written as:
\begin{gather}
	\label{eqn:Fpar}
	W_L = \sum_{j=1}^{N} \omega_{Lj}^2 = \frac{\sum_i(1 - \cos\Delta\phi_i)^2}{2\sum_i(1 - \cos\Delta\phi_i)},
\end{gather}
where we have used that $|\vec{Z}|^2= 2\sum_i(1 - \cos\Delta\phi_i)$. $W_L$ quantifies the non-linearity of a jump: the bigger $W_L$, the larger are the dominant $\Delta \phi_i$, and hence, the more non-linear is the jump. Fig.~\ref{fig:WL} shows the average of $W_L$ over avalanches as a function of system size, which seems to saturate  to a fairly large value of the order of $0.2$ in the thermodynamic limit. This is consistent with the findings of Fig.~\ref{fig:jump_densities}. Both show that large non-linear jumps are  frequent among the avalanche events. 

The weights of transverse modes are given by:
\begin{gather}
	\omega_{Tj} = -\dfrac{\sum_{i=1}^N\,a_i^{(j)}\,\sin\Delta\phi_i}{\sqrt{2\sum_{i=1}^N(1 - \cos\Delta\phi_i)}},
\end{gather}
where $a_i^{(j)} =\vec v_{Tj,i}$ is the $j$'th normalized eigenvector of the Hessian $T$~\eqref{eqn:HessianT}. In what follows, we focus on those modes only. They dominate the smaller jumps, which are only weakly non-linear.

\begin{figure}
	\includegraphics[width=0.95\columnwidth]{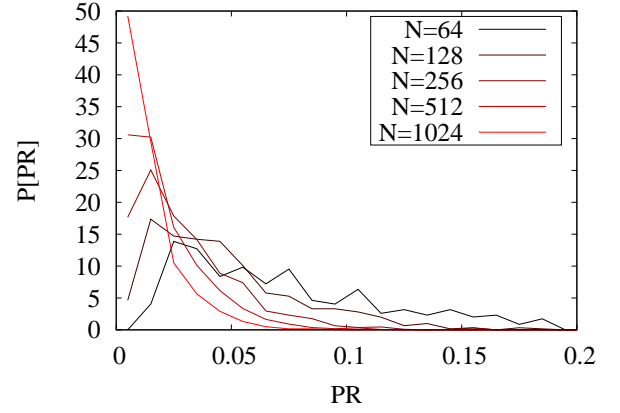}
	\includegraphics[width=0.95\columnwidth]{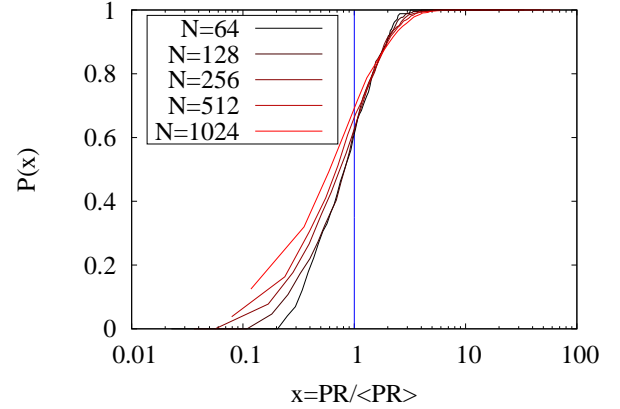}
	\caption{(Color online) \emph{Top}: Distribution of the participation ratio $Y_\omega$ of transverse modes, as defined in Eq.~\eqref{eqn:pr}. \emph{Bottom}: Cumulative distribution function of $Y_\omega$, rescaled by its average, for different system sizes. The absence of a clear collapse onto a single curve indicates the presence of many scales in the distribution of $Y_\omega$.}
	\label{fig:jumps_softest_mode}
\end{figure}



We define the participation ratio:
\begin{gather}
	\label{eqn:pr}
	Y_\omega = \frac{1}{N} \frac{\left[\sum_{j=1}^N \omega_{Tj}^2\right]^2}{\left[\sum_{j=1}^{N} \omega_{Tj}^4\right]},
\end{gather}
to characterize correlations between the linear modes of $T$ and the non-linear jump. $\tpr$ quantifies how many of the eigenmodes of $T$ contribute effectively to a jump.


The distribution of $Y_\omega$, shown in Fig.~\ref{fig:jumps_softest_mode} has a rather complex structure. In particular it does not exhibit a simple scaling with system size. Indeed, upon rescaling the cumulative distribution function (CDF) of $Y_\omega$ by the average, $\langle Y_\omega\rangle$, does not collapse the data for different system sizes. This indicates that jumps with different scalings are involved. This is also consistent with the scalings of various observables related to $Y_\omega$: The average $Y_\omega$ is found to scale like $N^{-0.59}$. The typical value i.e., the logarithmic average $\overline{\ln Y_\omega}$~\cite{parisi1987spin}, and the median $Y_\omega$ scale like $N^{-0.65}$, while the $10$th percentile (from the side of small participation ratios) scales like $N^{-0.75}$. These findings suggest that there are largely different jump events, small ones that one finds to be dominated by the softest modes of the susceptibility matrix, and large, strongly non-linear jumps, which have much less in common with the linear modes of the susceptibility matrix.

\section{Discussion and conclusion}
\label{sec:conclusions}

In this paper we have studied avalanche phenomena along the hysteresis loop in the fully-connected XY-spin-glass at zero temperature. Avalanches are triggered when the softest collective mode of the inverse susceptibility matrix becomes soft. This happens rather frequently, avalanches being separated only by increments $\delta H\sim N^{-\alpha}$ with $\alpha \approx 0.57 \pm 0.01$.  We observe that the softest modes of the inverse susceptibility matrix account for a large fraction of the non-linear avalanche events  for small jumps, similarly as in jammed soft matter systems. For big jumps, however, many more modes of the susceptibility matrix contribute.  

Let us now discuss a few of the interesting findings of this work. Interestingly, the soft modes triggering avalanches have a fractal support on the spins, involving only $\sim N^{1/3}$. This shows that the Hessians that occur in metastable minima of spin-glass problems are in fact non-trivially correlated random matrices, since in standard random matrix ensembles the eigenvectors have extensive participation ratios, rather than being fractals. So far, the understanding of the participation ratio $N^{1/3}$ is indirect, and based on a number of assumptions whose status is not fully clear. A more direct analytical understanding of properties of soft modes in spin-glass minima would thus definitely be of interest.   

The  sizes of magnetization jumps in avalanches extend up to scales set by the system size, similarly as in Ising spin-glasses. However, in contrast to the latter, the XY-glass is  found not to display self-organized criticality. Namely, avalanches \emph{typically} involve a finite fraction (of the order of $5\%$) of all spins, instead of being distributed according to a scale-free, decreasing power law. It would be interesting to understand whether this difference between spin-glasses with discrete and continuous degrees of freedom extends to other systems as well, and what are the mechanisms that lead to, or prevent, self-organized criticality. 

The absence of self-organized criticality, together with considerations about the similarity between off-equilibrium and equilibrium response, hints at the possibility that the ground state of the fully-connected XY-glass, and presumably of fully-connected vector glasses in general, might be described by a replica symmetry breaking order parameter function $q(x)$, which is not simply continuous as in the Ising case, but  might rather have discontinuous jumps in the low-$T$ limit, as well. To test this conjecture and to better understand the difference with the Ising case, it would therefore be interesting to find the $T\to 0$ limit of the equilibrium solution for these vector spin-glasses. 

We have found that the states visited along the hysteresis loop are actually not very strongly out-of-equilibrium. Indeed, the width of the full hysteresis loop is found to be subextensive, unlike in finite dimensional systems. A deeper analytical insight into why and how the $T=0$-dynamics remains so close to equilibrium is an interesting question for future studies of avalanche dynamics.

\begin{acknowledgments}
	We thank G. Biroli, S. Franz, L. Leuzzi, P. Young and M. Wyart for discussions. The numerical simulations were carried out with the aid of the Computer System of High Performance of the International Institute of Physics - UFRN, Natal, Brazil.
\end{acknowledgments}

\appendix

\section{Derivation of the inverse susceptibility matrix from finite $T$ TAP equations}
\label{app:TAP}

The aim of this appendix is to derive the results of Sec.~\ref{sec:model} starting from the Thouless-Anderson-Palmer (TAP) equations at finite temperature, and taking the limit $T\to 0$, so as to have full control over the Onsager back reaction (the last term in Eq.~\eqref{eqn:TAPeq} below). The TAP equations for vector spin-glasses were derived by Bray and Moore in Ref.~\onlinecite{bray1981metastable}:
\bea
	\label{eqn:TAPeq}
	\vy_i &=& H\vex + \sum_j J_{ij}\vm_j - \frac{\beta}{2}(1-q)\vm_i\\
	\label{eqn:eqn_m}
	\vm_i &=& \frac{\vy_i}{|\vy_i|}L(\beta |\vy_i|),
\eea
c.f., their equations ($4.9$, $4.10$). $L(x)$ is the Langevin function for XY spins,
\begin{gather*}
	L(x) = I_1(x)/I_0(x),
\end{gather*}
with $I_{0,1}(x)$ being modified Bessel functions. The overlap $q$ is defined by
\begin{gather*}
	q = \frac{1}{N}\sum_i\vm_i\cdot\vm_i.
\end{gather*}

Note that we use a different inverse temperature scale as compared to Ref.~\onlinecite{bray1981metastable}, $2\beta_\text{BM} = \beta$. The "thermodynamic field" $\vy_i$ appearing here is related to the average field $\vh_i$ defined in Eq.~(\ref{eqn:def-local-field}), by the Onsager shift:
\begin{gather}
	\vh_i = \vy_i + \frac{\beta}{2}(1-q)\vm_i.
\end{gather}

There are two equivalent ways to proceed in order to take the $\beta\to\infty$ limit. Let us first analyze the magnetic response to a homogeneous field at finite $T$, 
\begin{gather*}
	\vchi_i = \frac{\partial\vm_i}{\partial H} = \frac{\partial}{\partial H}\left(\frac{\vy_i}{|\vy_i|}\right) L(\beta |\vy_i|) + \frac{\vy_i}{|\vy_i|}\frac{\partial L(\beta |\vy_i|)}{\partial H},
\end{gather*}
and only then take the $T\to 0$ limit. We will see below that this limit commutes with the differentiation, however. 

Differentiation of the TAP equations~\eqref{eqn:TAPeq} yields, upon using the definition of $q$,
\begin{gather*}
	\frac{\partial\vy_i}{\partial H} = \vex + \sum_j J_{ij}\vchi_j - \frac{\beta}{2}(1-q)\vchi_i + \frac{\beta}{N}\left(\sum_i\vchi_i\cdot\vm_i \right)\vm_i.
\end{gather*}
and 
\begin{gather*}
	\frac{\partial}{\partial H}\left(\frac{\vy_i}{|\vy_i|}\right) = \frac{P^i}{|\vy_i|}\left(\frac{\partial \vy_i}{\partial H}\right)\notag\\
	= \frac{P^i}{|\vy_i|}\left(\vec{e}_{x} + \sum_j J_{ij}\vchi_j - \frac{\beta}{2}(1 - q)\vchi_i\right)\\
\end{gather*}
where the $2\times 2$-matrix $P^i$ projects on the component transverse to $\vm_i$, as defined after Eq.~\eqref{eqn:eqn_m}. With this, we obtain the expression for the susceptibility $\vchi_i$ at arbitrary temperature:
\begin{gather}
	\vchi_i = \frac{P^i}{|\vy_i|}\left(\vec{e}_{x} + \sum_j J_{ij}\vchi_j - \frac{\beta}{2}(1 - q)\vchi_i\right) L(\beta |\vy_i|)\notag\\
	\label{eqn:penult}
	 + \frac{\vy_i}{|\vy_i|}\frac{\partial L(\beta |\vy_i|)}{\partial H}.
\end{gather}
Projecting with $\vm_i$ from Eq.~\eqref{eqn:eqn_m} we obtain
\begin{gather*}
	\vchi_i \cdot \vm_i = L(\beta |\vy_i|)L^\prime(\beta |\vy_i|)\frac{\beta}{|\vy_i|}\left(\frac{\partial\vy_i}{\partial H}\cdot\vy_i\right).
\end{gather*}
One verifies that the function $x\,L(x)\,L^\prime(x)$ tends to zero as $x\to\infty$, which is a consequence of the fact that at $T=0$ the magnetic field cannot change the magnitude of the magnetization $|\vm_i|=1$. Therefore $\vchi_i \cdot \vm_i $, as well as the second term in~\eqref{eqn:penult}, vanish as $T\to 0$.

\subsection*{Onsager term}

Next we analyze the term $\beta(1-q)$:
\bea
	\frac{\beta}{2}(1 - q) &=& \frac{\beta}{2}\left[1 - \frac{1}{N}\sum_i\,L^2(\beta |\vy_i|)\right]\notag\\
	&\to & \frac{1}{2N}\sum_i\frac{1}{|\vy_i|} = \frac{1}{2h_\text{HM}},
\eea
since $L(x) = 1-1/2x$ as $x\to\infty$. Here, $h_\text{HM}$ is the harmonic mean of the fields $|\vy_i|$ over all the sites. 

Inserting this in Eq.~\eqref{eqn:penult} we obtain the susceptibility in the zero temperature limit:
\begin{gather}
	\label{eqn:chi}
	\vchi_{i\mu} = \frac{P^i_{\mu\nu}}{|\vy_i|}\left[\delta_{\nu,x} + \sum_j J_{ij}\chi_{j\nu} - \frac{\vchi_{i\nu}}{2h_\text{HM}}\right]\,.
\end{gather}
Therefore the susceptibility is a solution of the following matrix equation ($\mu,\nu$ refer to spin components):
\bea
	\label{eqn:mat}
	&&\sum_{j\nu} A_{i\mu,j\nu} \chi_{j\nu} = C_{i\mu},\\
	&&A_{i\mu,j\nu} = |\vy_i|\delta_{ij}\delta_{\mu\nu} + \frac{1}{2h_\text{HM}}\delta_{ij}P^i_{\mu\nu} - J_{ij}\sum_\sigma P^i_{\mu\sigma}P_{\sigma\nu}^j,\notag\\
	&&C_{i\mu} = \delta_{\mu x} - m_{i\mu}m_{i x}.\notag
\eea
This expression is almost identical to Eq.~\eqref{eqn:defA_mx} except for the extra term proportional to $1/2h_\text{HM}$. However, it is immediate to see that in the transverse sector the susceptibility matrix, $T$~\eqref{eqn:defT_mx}, is exactly the same. Projecting from both sides with  $\vn_i = (m_{iy},-m_{ix})$, we have:
\begin{gather}
	T_{ij} = n_{i\mu}A_{i\mu,j\nu}n_{j\nu} = \left(|\vy_i| + \frac{1}{2h_\text{HM}}\right)\delta_{ij} - J_{ij} \vm_i\cdot\vm_j,
\end{gather}
while in the longitudinal sector we find
\begin{gather}
	m_{i\mu}A_{i\mu,j\nu}m_{j\nu} = |\vy_i|\delta_{ij}.
\end{gather}
At $T=0$ we have $\vy_i\parallel\vm_i$ and $|\vh_i| = |\vy_i| + 1/2h_\text{HM}$. The only effect of the Onsager back reaction is to modify the longitudinal spectrum of the inverse susceptibility matrix, replacing average fields $\vh_i$ by thermodynamic fields $\vy_i$. The relation $|\vh_i| = |\vy_i| + 1/2h_\text{HM}$ immediately implies a hard gap in the distribution of $|h_i|$ of at least $1/2h_\text{HM}$ as shown in Fig~\ref{fig:loc-fields-dist}. This lower bound on the hard gap is expected to be tight~\cite{bray1981metastable}.


\subsection*{Direct $T=0$ limit}

The above result can also be obtained by setting $T=0$ directly within the TAP equations~(\ref{eqn:TAPeq}-\ref{eqn:eqn_m}), and differentiating afterwards. Since $\beta(1-q) \to \frac{1}{h_\text{HM}}$ and $L(\beta \vy_i)\to 1$, the TAP equations become
\begin{gather*}
	\vy_i = H\,\vex + \sum_j J_{ij}\vm_j - \frac{1}{2h_\text{HM}}\vm_i,\\
	\vm_i = \frac{\vy_i}{|\vy_i|}.
\end{gather*}
Eq.~\eqref{eqn:mat} follows from this by differentiation with respect to $y_i$. \\


\section{Inverse susceptibility matrix $T_{ij}$ as the Hessian of the angular energy functional $\mathcal{H}(\phi)$}
\label{Tmatrix:deriv}

In this appendix we demonstrate that the transverse inverse susceptibility matrix $T_{ij}$~\eqref{eqn:defT_mx} follows naturally from the angular energy functional~\eqref{eqn:ham_phi}. We again neglect the Onsager term.
Let us analyze directly the angular energy functional of Eq.~\eqref{eqn:ham_phi}:
\begin{gather}
	\mathcal{H} = -\frac{1}{2}\sum_{ij}J_{ij}\cos(\phi_i-\phi_j) - H\sum_i\cos(\phi_i),
\end{gather}
and establish its relationship with the $T\to 0$ limit of the TAP equations.

Indeed, its Hessian is
\bea
	\frac{\partial^2\mathcal{H}}{\partial\phi_i\partial\phi_j} &=& \delta_{ij}\Big[H\cos(\phi_i) + \sum_k J_{ik}\cos(\phi_i - \phi_k)\Big] \nn\\
	& &- J_{ij}\cos(\phi_i - \phi_j).
	\label{eqn:Hessian}
\eea

Recalling the definition of average fields, Eq.~(\ref{eqn:def-local-field}),
\begin{gather*}
	\vh_i = H\vex + \sum_k J_{ik}\vm_k,
\end{gather*}
one easily sees that the coefficient of $\delta_{ij}$ in \eqref{eqn:Hessian}  is the projection of $\vh_i$ onto the unit vector $\vm_i = \vh_i/|\vh_i|$, i.e. $\vh_i\cdot\vm_i = |\vh_i|$. This establishes the equivalence of the Hessian~\eqref{eqn:Hessian} with the transverse inverse susceptibility matrix $T_{ij}$ of Eq.~\eqref{eqn:defT_mx},
\begin{gather}
	\frac{\partial^2\mathcal{H}}{\partial\phi_i\partial\phi_j}  = T_{ij}.
\end{gather}

\bibliography{xy_jumps}{}

\end{document}